\documentclass[a4paper]{article}
\usepackage{amssymb}
\usepackage{graphicx}
\usepackage{amssymb}
\usepackage{amsmath}
\usepackage{xcolor}
\usepackage[cm]{fullpage}
\usepackage[numbers]{natbib}
\usepackage[hidelinks]{hyperref}

\newcommand{\figureref}[1]{Figure~#1}
\newcommand{\figuresref}[1]{Figures~#1}
\newcommand{\suppref}[1]{Supplementary Information Section~#1}

\begin{document}
	
	\title{Coherent shuttle of electron-spin states}
	\author{T. Fujita$^{1}$, T.~A. Baart$^{1}$,\\ C. Reichl$^{2}$, W. Wegscheider$^{2}$, L.~M.~K. Vandersypen$^{1}$\footnote{email: l.m.k.vandersypen@tudelft.nl}}
	\maketitle
	
	\begin{enumerate}
		\item QuTech and Kavli Institute of Nanoscience, TU Delft, 2600 GA Delft, The Netherlands
		\item Solid State Physics Laboratory, ETH Z\"{u}rich, 8093 Z\"{u}rich, Switzerland
	\end{enumerate}
	
	\section*{Abstract}
	We demonstrate a coherent spin shuttle through a GaAs/AlGaAs quadruple-quantum-dot array. Starting with two electrons in a spin-singlet state in the first dot, we shuttle one electron over to either the second, third or fourth dot. We observe that the separated spin-singlet evolves periodically into the $m=0$ spin-triplet and back before it dephases due to nuclear spin noise. We attribute the time evolution to differences in the local Zeeman splitting between the respective dots. With the help of numerical simulations, we analyse and discuss the visibility of the singlet-triplet oscillations and connect it to the requirements for coherent spin shuttling in terms of the inter-dot tunnel coupling strength and rise time of the pulses. The distribution of entangled spin pairs through tunnel coupled structures may be of great utility for connecting distant qubit registers on a chip.
	
	\section*{Introduction}
	Single electron spins in semiconductor quantum dots have been proposed as a candidate qubit platform that may allow scalable quantum information processing~\cite{Loss1998a,Taylor2005}. Recent progress in semiconductor quantum-dot structures have shown long single-spin coherence times and high-fidelity coherent operations~\cite{Petta2005,Nowack2007,Hanson2007,Kawakami2014,Veldhorst2015}. Furthermore, linear arrays have been successfully scaled to triple and quadruple dots~\cite{Medford2013,Eng2015,Baart2016, Takakura2014,Baart2016a}, and 2$\times $2 arrays have been demonstrated as well~\cite{Thalineau2012}. However, there are practical limitations to the size of tunnel-coupled quantum dot arrays in one or two dimensions. Integrating larger numbers of qubits can be achieved by coherently connecting distant qubit registers on a chip~\cite{Childress2004, Trifunovic2012, Hassler2015,Schuetz2015}. Such coherent links could also serve to connect different functions such as memory and processor units~\cite{Taylor2005}. As an alternative to coherent spin-spin coupling at a distance, the physical transfer of electrons across the chip while preserving the spin information can serve as an interface between separated quantum dot arrays~\cite{Taylor2005,Oskin2003}. This is similar in spirit to experiments with trapped ions that were shuttled around through segmented ion traps~\cite{Kielpinski2002,Nakajima2016}. To our knowledge, there are no demonstrations of the transfer of single electron spins coherently through multiple quantum dots.\\
	
	Various physical mechanisms have been proposed to controllably transfer single charges through confined structures, including Thouless pumps~\cite{Gefen1987,Aleiner1998}, charge pumps~\cite{Pothier1992}, and surface acoustic waves~\cite{Shilton1996}, and several of these approaches have been experimentally demonstrated with quantum dot devices~\cite{Baart2016a,Hermelin2011,McNeil2011}. Furthermore, charge transfer with preservation of spin projection was shown using surface acoustic waves~\cite{Bertrand2015a} and charge pumps~\cite{Baart2015}. However, none of these approaches has proceeded to measure the preservation of spin coherence during the shuttling processes. The main obstacle towards spin coherent transfer is the short dephasing time due to nuclear spins in GaAs, the commonly used host material for shuttling experiments so far ($T_{2}^{*}$ of 10-30 ns)~\cite{Chekhovich2013}.\\
	
	In this Article we explore the coherent shuttling of single electron spins through a quadruple-quantum-dot array by applying gate-voltage pulses to push the electron through the array. Starting with two electrons in a spin-singlet state in a single dot at one extremity, we shuttle one of these two electrons to the second, third and fourth dot. We probe the preservation of spin coherence by subsequently attempting to bring the two electrons back onto the first dot. Through the Pauli exclusion principle, this is allowed only if the spin singlet state is still preserved. Starting from an analysis of the spin-shuttle performance, we discuss two adiabaticity conditions of the transfer, one for coherent single-spin shuttling and one for distributing entangled states.
	
	\section*{Methods and Results}
	
	The quadruple-quantum-dot array is formed electrostatically in a two-dimensional electron gas (2DEG) 90~nm below the surface of a GaAs/AlGaAs heterostructure, see \figureref{1a}. Gate electrodes fabricated on the surface are biased with appropriate voltages to selectively deplete regions of the 2DEG and define the linear array of four quantum dots, designated as dot~1, 2, 3, and 4 counting from the left. A sensing dot (SD1) next to the quantum dot array is used for non-invasive charge sensing using radiofrequency (RF) reflectometry to monitor the number of electrons in each dot~\cite{Barthel2010}. We denote the charge state occupation of the array as $(nmpq)$ referring to  the number of electrons in dot~1, 2, 3, and 4 respectively. We tune the quadruple dot such that it is occupied with two electrons in total. Together the two electrons can form a spin-singlet state $S$, or one of the three triplet states $T_{0}$, $T_{-}$, and $T_{+}$. An in-plane magnetic field $B_{\rm ext}$ = 2.3~T is applied to split the $T_{-}$ and $T_{+}$ states from the $T_{0}$ state. In order to shuttle electrons back and forth through the array within the dephasing time $T_{2}^{*}$ of a few tens of ns, all inter-dot tunnel couplings are tuned to about 1~GHz and higher (see \suppref{III} for the procedure). The device is cooled inside a dilution refrigerator to a base temperature of ${\sim}22$~mK (see \suppref{I} for more details on the measurement setup).\\
	
	To study the preservation of spin coherence while shuttling electrons around, we initialize the two electrons in a spin-singlet state in the leftmost dot, then move one electron over to either the second, third or fourth dot and back, and probe whether the spin-singlet phase is still preserved after shuttling. Spin-singlet initialization is done by waiting for thermal equilibration in the (2000) charge state (point I in \figureref{1b}). Next, one of the electrons is shuttled to one of the other three dots by ramping the voltages on gates P1 and P4 from position I to position A, B or C in \figureref{1b} in 2.5 ns. The ramp is chosen such that the electron motion is adiabatic with respect to the inter-dot tunnel coupling to ensure that the electron moves to the neighbouring dots in a reproducible manner, see \figureref{2a}. Pulsing too fast would make the electron lag behind and tunnel inelastically at a random later time. As we will show below, the spin precession rates in the four dots differ from each other by tens of MHz, and therefore tunnelling at a random time would contribute to a loss of spin coherence. At the same time, the transfer should be completed well within the $T_2^*$. We will introduce an additional timing constraint in the discussion section. Furthermore, by tuning the dot-reservoir couplings to be much smaller ($\sim$~kHz) than the inter-dot tunnel couplings, the shuttling of an electron to any of the dots can be implemented in one step - as opposed to dot by dot - without a significant probability of losing the electron to the reservoirs when passing through the (1000) region, see \figureref{2b}. After the forward shuttle, this electron is pulled back towards dot~1 for a spin measurement (point M in \figureref{1b}). If the spin state remains a singlet, the two electrons can both reside in dot~1 and the charge sensor will indicate charge state (2000). If the spin-state has changed to a triplet, the second electron is stuck on dot~2 due to Pauli-spin-blockade (PSB)~\cite{Ono2002}and the charge sensor will record (1100). We calibrated the read-out fidelities to be over 95~\% for both the spin-triplet and spin-singlet (see \suppref{IV}). Through this method of spin-to-charge conversion we can tell whether the phase between the two electron spins has changed during shuttling.\\ 
	
	The main experimental observations are as follows. When shuttling one of the electrons to dot~2 and varying the waiting time in that dot (point A in \figureref{1b}), we observe that the singlet-return probability oscillates sinusoidally with a frequency of $103 \pm 5$~MHz (see \figureref{3a}). When shuttling to dot~3, a similar oscillation is observed (see \figureref{3b}), albeit at a frequency of $118 \pm 7$~MHz. For dot~4 we record an oscillation with a frequency of $185 \pm 14$~MHz (see \figureref{3c}). The oscillations have a contrast of $0.63 \pm 0.12$, $0.62 \pm 0.14$, and $0.57 \pm 0.15$ and decay on a timescale of $14 \pm 3$, $13 \pm 3$ and $7 \pm 2$~ns respectively. 
	
	\section*{Discussion}
	The observed oscillations can be interpreted in terms of qubit evolutions between the singlet state, $S$, and the $m=0$ triplet state, $T_{0}$, around two orthogonal axes governed by the exchange term, $J \vec{\sigma_i}\cdot\vec{\sigma_j}/4$, and a local difference in Zeeman energy, $\Delta E_z^{i,j} (\sigma_i^z - \sigma_j^z)/2$, where $\vec{\sigma_i}$ is the Pauli matrix for the spin in dot $i$. These two axes are depicted in the Bloch sphere of \figureref{3d}. The magnitude of $J$ depends on the overlap of the two electron wavefunctions. Inside the (2000) region, $J$ is of order 1~meV with $S$ the ground state, allowing us to initialize in $S$. As we move from (2000) towards (1100), $J$ gradually decreases up to the point where $\Delta E_z^{i,j}$ dominates. For (1010) and (1001), $J$ is negligibly small. A difference in Zeeman splittings could result from a difference in local magnetic fields~\cite{Tokura2006, Foletti2009}. In the present experiment, the local fields are identical but the $g$-factors are different between the dots~\cite{Baart2015}. In this case, and as observed, $\Delta E_z^{i,j}$ increases linearly with $B_{\rm ext}$. Then, starting with a (2000) singlet, separating the electrons across two dots kick-starts a coherent $S$-$T_0$ evolution around $\Delta E_z^{i,j}$~\cite{Foletti2009}. Since $\Delta E_z^{i,j}$ is different for different pairs of dots, the frequency of the $S$-$T_0$ oscillations is dependent on the location of the two electrons (see also Ref.~\citenum{Nakajima2016a}, where a three-electron state in a triple dot was adiabatically transferred from $\uparrow_1 S_{23}$ to $\uparrow_2 S_{13}$, where the subscripts refer to the dot number). The damping of the oscillation on a timescale of about 10 ns is due to the hyperfine interaction with the randomly oriented nuclear spins in the host material of the two dots~\cite{Petta2005}. \\
	
	To test this interpretation of the measurements in \figureref{3} in terms of $S$-$T_0$ oscillations, we have measured the Zeeman energy in each dot using electric-dipole-spin-resonance (EDSR) measurements~\cite{Shafiei2013}. For this aim we modified the gate voltage settings in order to implement single-spin read-out, using the single-spin CCD protocol of Ref.~\citenum{Baart2015}. Chirped microwave signals were applied to gate P3 to adiabatically invert a spin from its initial state $\uparrow$ to $\downarrow$.  The $\Delta E_z^{i,j}$ extracted from the individual spin resonance frequencies for magnetic fields 2.7  to 4.7 T are depicted in \figureref{4a}. Extrapolating to 2.3 Tesla using a fit through the data points, we observe that the sequence of the frequency differences corresponds to that obtained from the $S$-$T_0$ oscillations, but the absolute values differ. This could be due to the fact that the gate voltages used for the EDSR measurements were different from those used for the $S$-$T_0$ oscillations by 5-15~mV. In EDSR measurements on dot~1 and 4, we have found comparable changes in the spin splittings with similar gate voltage changes (see \suppref{V}). Within this offset, the three distinct oscillation frequencies in the data of \figuresref{3a,b,c} support the interpretation that one of the electrons was shuttled controllably from dot 1 to dot 2, 3 and 4, whilst preserving phase coherence.\\
	
	The oscillation contrast provides quantitative information on the spin-singlet component in the two-electron state after shuttling. In the experiment, the measured contrasts are well below unity. To investigate whether the reduced contrasts are due to dephasing upon shuttling or have a different origin, we numerically simulate the shuttling process. The simulations compute the spin evolution taking into account the finite rise time of the gate voltage pulse, the inter-dot tunnel couplings, the Zeeman energy differences and inelastic relaxation. Starting from a $S$(2000) state, we calculated the population of the various states after pulsing the dot levels in a way that resembles the shuttle experiments (see \suppref{VI} for the details of the simulation). For simplicity, we initially leave out inelastic relaxation as well as the return pulse used for spin read-out. Figures 4b,c,d show the numerically computed contrast for the $S$-$T_0$ oscillations after shuttling to the (1100), (1010), and (1001) charge regions, as a function of the inter-dot tunnel coupling. For the inter-dot transition from (2000) to (1100) the contrast is peaked around 1.2 GHz (see \figureref{4b}). The increasing contrast for small tunnel couplings, which appears also for the other inter-dot transitions, arises from the increasing probability that the charge is (adiabatically) transferred. Obviously, if both charges stay in the first dot, there will be no $S$-$T_0$ oscillations. If the charge initially stays behind in the first dot and then tunnels inelastically to a next dot at a (random) later time, the spin phase will be out of sync with the case of a adiabatic charge transfer. This also reduces the $S$-$T_0$ oscillation contrast. As tunnel couplings increase further, the transition from a Hamiltonian dominated by spin exchange to one dominated by the Zeeman energy difference becomes wider, which eventually results in a spin-adiabatic transition. In this case, the system adiabatically moves from $S$(2000) to $\downarrow\uparrow$(1100) and the amplitude of the $S$-$T_0$ oscillations gradually vanishes. When moving from (1100) to (1010) and (1001), the spin eigenbasis does not change and we can afford higher tunnel couplings without a rapid degradation of the contrast (see \figuresref{4c,d}). The gradual decreases in contrast seen at higher tunnel couplings occurs because the electron is increasingly delocalized over the strongly coupled dots, and the plots show the amplitude of the $T_0$ oscillations for a single charge state only. \\
	
	Before proceeding, we point out that the spin-adiabaticity condition that applies when starting from a (2000) singlet, does not apply when starting from (1000). In the latter case, the spin eigenstates are given by $\{\uparrow,\downarrow\}$ in any of the dots. We note that the spin adiabaticity condition is particular to separating a two-spin state and does not apply when shuttling a single-spin state across an array.\\
	
	Adding to the simulation the return path introduces additional phase shifts and offsets in the simulated $S$-$T_0$ oscillations versus waiting time, without a significant further effect on the contrast of the oscillations (see \suppref{VI}). Overall, the contrast in the simulated oscillations does not exceed 0.65-0.8. This is only about 10\% larger than the experimentally observed contrasts, which also incur measurement infidelities of 4-5\%.  This suggests that the tunnel coupling between dot~1 and dot~2 in the experiment was close to the optimal value of around 1.2~GHz. Importantly, the contrast of the $S$-$T_0$ oscillations being only 60-65\% does not imply a loss of phase coherence by 30-35\%, but is mostly due to the difficulty in satisfying competing adiabaticity conditions. If the pulse rise time is too fast, the charge transition is non-adiabatic (the charge does not follow the pulse but stays in (2000) all along). If the pulse rise time is too slow, the spin transition is adiabatic (the singlet is rotated to $\downarrow\uparrow$). In both cases, the (2000) spin singlet is recovered after pulsing back. Only the case of inelastic tunnelling at a random time contributes to the loss of phase coherence, by an amount that mostly depends on the inelastic tunnel rate versus the $S$-$T_0$ oscillation frequency.\\
	
	From comparing the measured and simulated contrasts as a function of the shuttling distance, we conclude that phase coherence is largely preserved when shuttling spins through the array. Apart from inelastic tunnelling at a random time, the relevant mechanisms that could compromise phase coherence during shuttling are the spin-orbit interaction and hyperfine interaction. Spin-orbit interaction would induce a deterministic spin rotation as an electron moves across the array~\cite{Schreiber2011}. In the present experiment, the effect of spin-orbit interaction is minimized for motion along the inter-dot axis by the choice of alignment of the gate design with respect to the crystallographic axes and the external magnetic field (see \figureref{1a})~\cite{Baart2016}. The slowly varying hyperfine interaction gives rise to random nuclear fields in each of the quantum dots for every shuttle run. In our experiment the nuclear spin distribution specific to the final charge state contributed to the damping of the $S$-$T_0$ oscillations. However, when shuttling rapidly through large dot arrays, the effect of random nuclear fields increasingly averages away, and the spin is preserved better than if the electron stayed in a single charge state~\cite{Echeverria-Arrondo2013}. Both hyperfine and spin-orbit interaction are significantly weaker in silicon than in GaAs, so spin shuttles in silicon should be even more robust. 
	
	\section*{Conclusion}
	We demonstrate a coherent spin-shuttle though a quadruple-quantum-dot device. The main observation is a coherent singlet-triplet oscillation that occurs when one electron of a spin-singlet pair is shuttled to a distant dot. Such an oscillation can only be observed if the coherence is preserved during shuttling. The key requirement for a coherent spin shuttle in the presence of differences in Zeeman splitting between the dots in the array, is that the motion through the array be adiabatic with respect to the inter-dot tunnel couplings for the spin phase to be preserved. For distributing entangled states such as the spin singlet, an additional condition is that the transitions must be non-adiabatic with respect to the spin exchange strength. This demonstration and analysis open up new avenues for large scale solid-state quantum computation, whereby single electrons are shuttled and entanglement is distributed across the chip~\cite{Saraga2003,Taylor2005a}.

	\newpage
	
	\subsection*{Acknowledgements} The authors acknowledge useful discussions with the members of the Delft spin qubit team and J.M. Taylor and experimental assistance from M. Ammerlaan, J. Haanstra, R. Roeleveld, R. Schouten, M. Tiggelman and R. Vermeulen. This work is supported by the Netherlands Organization of Scientific Research (NWO) Graduate Program, the Intelligence Advanced Research Projects Activity (IARPA) Multi-Qubit Coherent Operations (MQCO) Program, the Japan Society for the Promotion of Science (JSPS) Postdoctoral Fellowship for Research Abroad, and the Swiss National Science Foundation.
	
	\subsection*{Author contributions}
	T.F. and T.A.B. performed the experiment and analysed the data, C.R. and W.W. grew the heterostructure, T.F., T.A.B., and L.M.K.V. contributed to the interpretation of the data, and T.F., T.A.B. and L.M.K.V. wrote the manuscript.

	\subsection*{Author information}
	The authors declare no competing financial interests. Correspondence and requests for materials should be addressed to L.M.K.V.
	
	\newpage
	
	\noindent \textbf{Figure 1 Device image and spin shuttling protocol}\\
	\textbf{a} Scanning electron microscope image of a sample nominally identical to the one used for the measurements. Dotted circles indicate the intended quantum dot positions and squares indicate Fermi reservoirs in the 2DEG, which are connected to ohmic contacts. The RF reflectance of SD1 and/or SD2 are monitored in order to determine the occupancies of the four dots in the linear array. \textbf{b} Charge stability diagram that includes the regions with two electrons involved in the shuttling sequence. Horizontal traces are averaged over 200 sweeps with a scan rate of 4.4~ms per sweep. The (1010) (or (1100)) region fading into the (1000) region occurs due to the slow unloading of the electron in dot~3 (or dot~2) relative to the scan rate. Labels indicate the number of electrons on each of the four dots. The dot array is initialized with two electrons on the leftmost dot, allowing them to relax to the spin-singlet ground state (point I). Then, a gate voltage pulse takes the system to point A, B or C, resulting in the transfer of one of the two electrons to the second, third or fourth dot, respectively. After the gate voltage pulse, the system returns to point (M), and the spin state is measured using Pauli spin blockade. We infer a spin-singlet (triplet) state when the signal of SD1 reads (2000) ((1100)) for an integration time of 5~$\mu$s. The data shown here was chosen since it clearly shows all the relevant charge transitions. The corresponding data with the settings used in \figureref{3} is shown in \suppref{II}.\\\\
	
	\noindent \textbf{Figure 2 Adiabaticity requirements for shuttling}\\
	\textbf{a} Schematic energy diagram for shuttling from (2000) (I) to (1100) (A). The rise time of the applied gate voltage pulse is chosen such that the charge motion is adiabatic with respect to the inter-dot tunnel coupling, causing one of the electrons to move to dot~2, but non-adiabatic with respect to the spin Hamiltonian causing the spin singlet state to be preserved. As a consequence, the system is taken from $S$(2000) to $\frac{\uparrow\downarrow - \downarrow\uparrow}{\sqrt{2}}$(1100) (orange dashed arrow). \textbf{b} A similar energy diagram when linearly moving between points I and C in \figureref{1b} (\suppref{II}). Lines are depicted for spin-less states. When inter-dot tunnel rates are tuned to be larger than the rate of change of the detuning (controlled by the ramp rate of $V_{\rm P1}-V_{\rm P4}$) and the dot-reservoir tunnel rates, the charge state adiabatically follows its original energy branch without crossing over to other branches and without the electron escaping to the reservoir (the horizontal dashed line indicates the Fermi-level). Furthermore, as in panel (a), the detuning ramp rate must be fast enough such that the spin singlet is preserved during the charge transfer.\\\\
	
	\noindent \textbf{Figure 3 Coherent shuttling and $S$-$T_0$ oscillations}\\
	\textbf{a-c} Measured spin-singlet probability as a function of the waiting time in the (1100), (1010), or (1001) charge state (error bars indicate 95~\% confidence intervals). This waiting time occurs in between separating the electrons and trying to bring them back together. The spin-singlet probability is fitted with a Gaussian damped cosine function with a constant phase shift. The fitted $S$-$T_0$ oscillation frequencies are: 103 $\pm 5$~MHz, 118 $\pm 7$~MHz, and 185 $\pm 14$~MHz respectively. The fitted contrasts are $0.63 \pm 0.12$, $0.62 \pm 0.14$, and $0.57 \pm 0.15$. The damping is attributed to the random nuclear fields in the two dots and the phase shift accounts for the additional shuttle towards the measurement point.. \textbf{d} Bloch sphere representation of a singlet-triplet qubit. For a spin singlet distributed over two separated dots, the exchange interaction $J$ is small and the Zeeman energy difference between the dots, $\Delta E_z^{i,j}$, dominates.\\\\
	
	\noindent \textbf{Figure 4 Zeeman energy differences and simulated $S$-$T_0$ oscillation contrasts }\\
	\textbf{a} The Zeeman energy in each dot is determined using EDSR measurements for a range of magnetic field strengths~\cite{Shafiei2013,Baart2015} and is fitted to $E_z=\left|g_1\mu_{\rm B}B +g_3\mu_{\rm B}B^3\right|$, resulting in $g_{1}^{\rm{dot1}}=-0.4332\pm 0.0005$, $g_{3}^{\rm{dot1}}=\left( 0.47\pm 0.03\right) \cdot10^{-3}$/T$^2$, $g_{1}^{\rm{dot2}}=-0.4353\pm 0.0004$, $g_{3}^{\rm{dot2}}=\left( 0.46\pm 0.04\right) \cdot10^{-3}$/T$^2$, $g_{1}^{\rm{dot3}}=-0.4361\pm 0.0006$, $g_{3}^{\rm{dot3}}=\left( 0.47\pm 0.04\right) \cdot10^{-3}$/T$^2$, $g_{1}^{\rm{dot4}}=-0.4372\pm 0.0004$, $g_{3}^{\rm{dot4}}=\left( 0.47\pm 0.03\right) \cdot10^{-3}$/T$^2$. The solid data points represent the differences in Zeeman energy between dot $n$ ($n = 2,3,4$) and dot 1 as a function of magnetic field, extracted from the EDSR measurements (error bars account for the expected fluctuations of $\pm$15~MHz from the quasi-static nuclear field and $\pm$5~MHz from the frequency resolution of the measurement, for each of the two resonance measurements used in the frequency difference calculation). Solid curves represent the difference between the fit of dot $n$ and dot 1. Open circles show the Zeeman energy differences extracted from \figureref{3a,b,c} (error bars indicate the fitting uncertainty). \textbf{b} The contrast of the simulated coherent oscillation when pulsing from point I to point A in \figureref{1b}. The simulation assumes a linear ramp from start to end over 2.50~ns. \textbf{c} Similar to (b) but pulsing to point B and varying the tunnel coupling between dots 2 and 3. \textbf{d} Similar to (b) but pulsing to point C and varying the tunnel coupling between dots 3 and 4. The tunnel couplings were set to $t_{12}=1.2$~GHz, $t_{23}=2.6$~GHz, and $t_{34}=4.3$~GHz if not varied. These values correspond to the peak positions in the tunnel coupling dependencies. Note that the maximum in {\bf d} is lower and occurs for a higher tunnel coupling than that in {\bf c} because the ramp rate is higher for {\bf d} (same ramp duration but higher pulse amplitude). 
	
	\newpage
	
	\begin{figure*}[h!]
		\centering
		\includegraphics[width=1.0\textwidth]{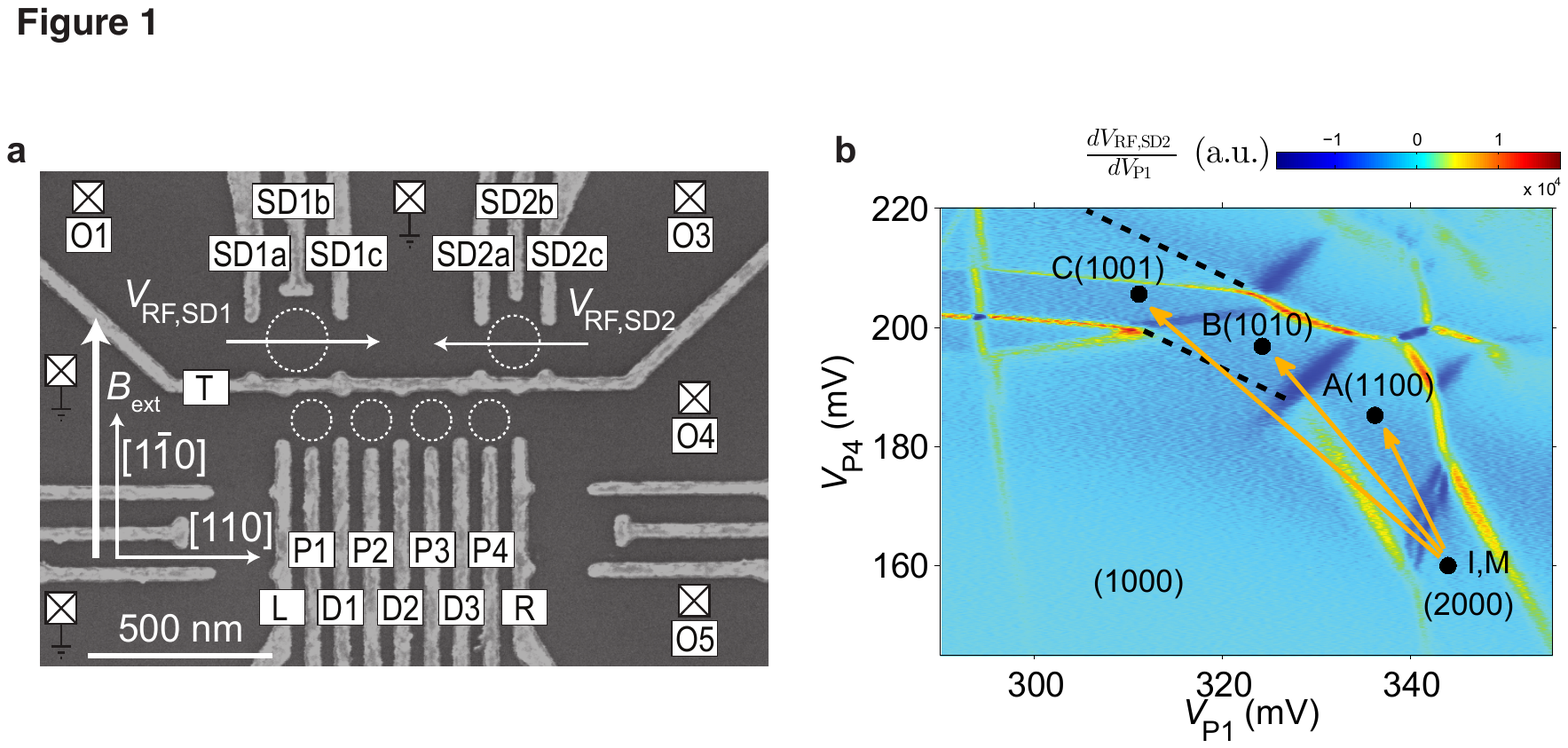}
	\end{figure*}
	
	\newpage
	
	\begin{figure*}[h!]
		\centering
		\includegraphics[width=1.0\textwidth]{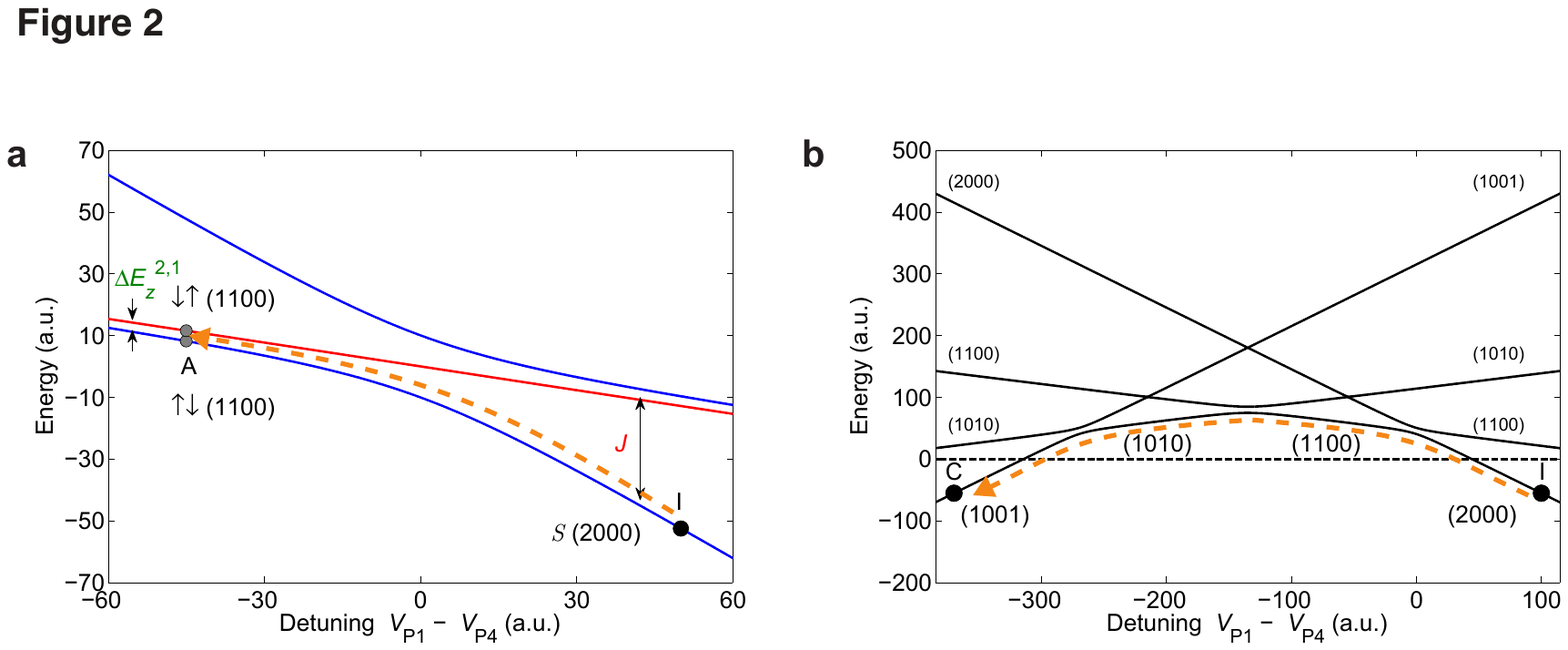}
	\end{figure*}
	
	\newpage
	
	\begin{figure*}[h!]
		\centering
		\includegraphics[width=1.0\textwidth]{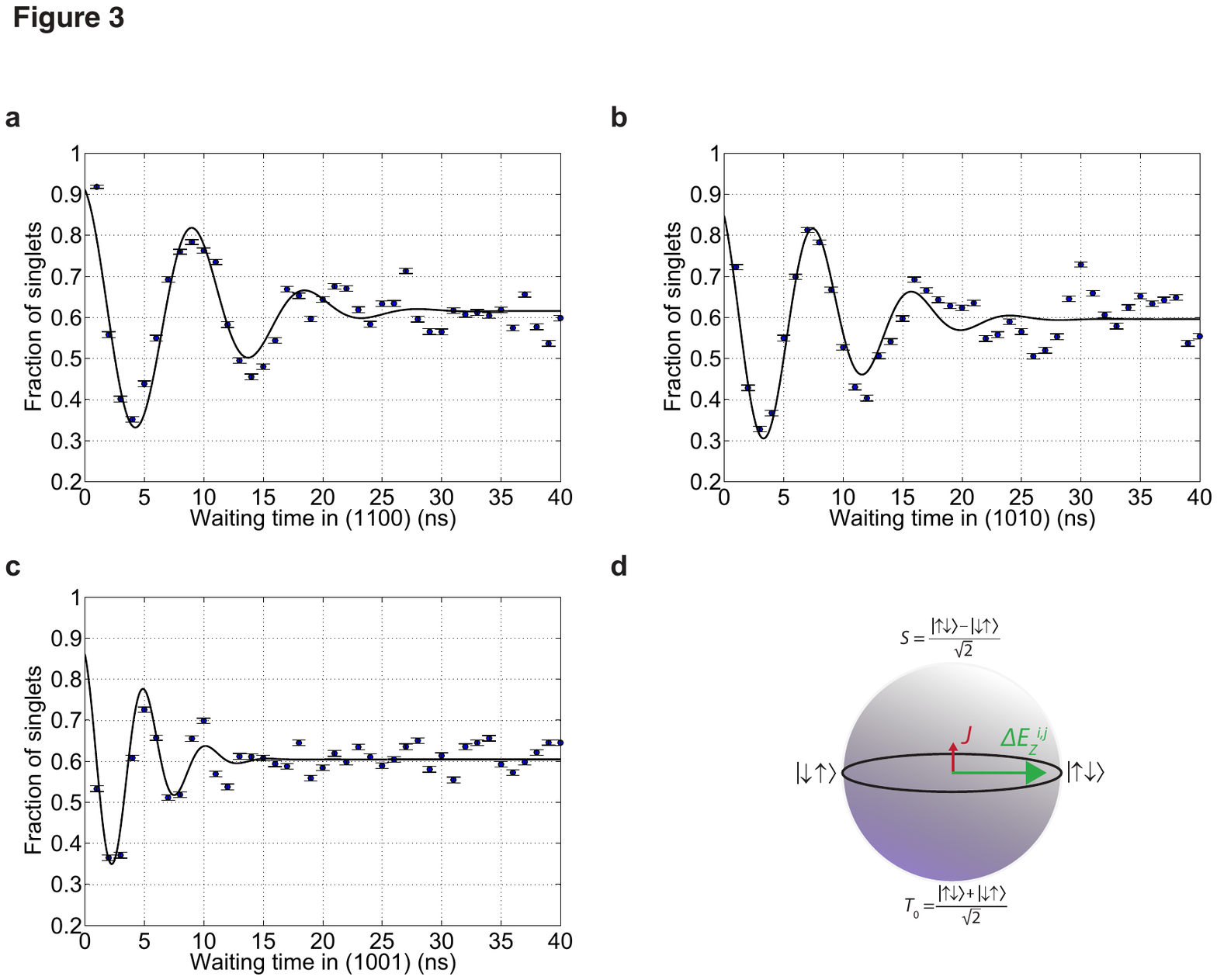}
	\end{figure*}
	
	\newpage
	
	\begin{figure*}[h!]
		\centering
		\includegraphics[width=1.0\textwidth]{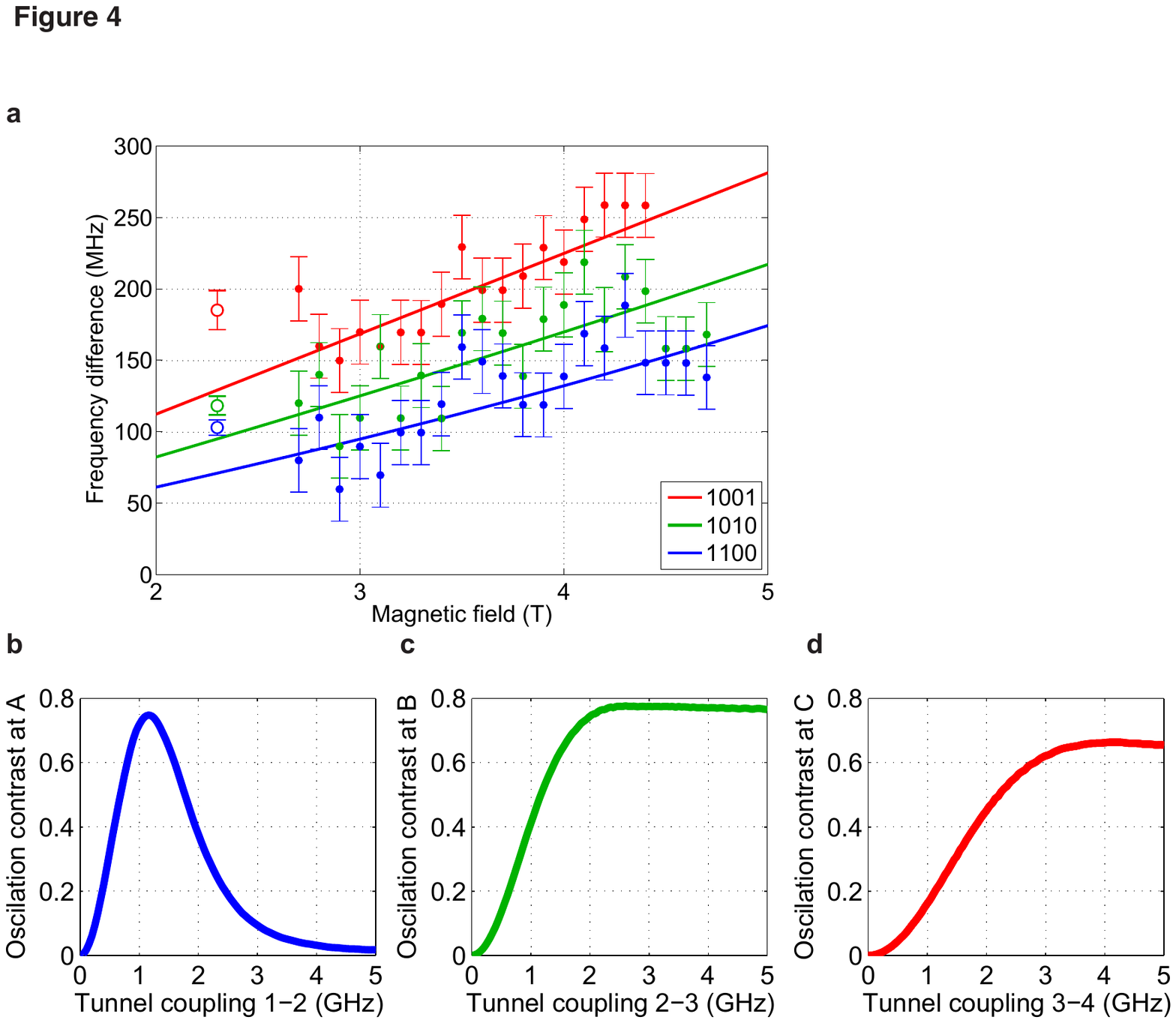}
	\end{figure*}
	
	\clearpage
	\newpage
	
	\renewcommand{\figurename}{Fig.~S}
	\renewcommand{\tablename}{Table~S}
	\renewcommand{\theequation}{S\arabic{equation}}
	\renewcommand{\thesection}{\Roman{section}.} 
	\renewcommand{\thesubsection}{\Alph{subsection}.}
	
	\begin{centering}
		{\Large Supplementary Information for} \\ \vspace{0.2cm}
		{\Large \textbf{Coherent shuttle of electron-spin states}}\\
		\vspace{0.4cm}
		{\normalsize T. Fujita$^{1}$, T.~A. Baart$^{1}$,}\\  
		{\normalsize C. Reichl$^{2}$, W. Wegscheider$^{2}$, L.~M.~K. Vandersypen$^{1}$}\\
		
		\vspace{0.4cm}
		\normalsize{$^{1}$QuTech and Kavli Institute of Nanoscience, TU Delft, 2600 GA Delft, The Netherlands}\\
		\normalsize{$^{2}$Solid State Physics Laboratory, ETH Z\"{u}rich, 8093 Z\"{u}rich, Switzerland}\\
	\end{centering}
	
	\renewcommand{\baselinestretch}{2.0}\normalsize
	\tableofcontents
	\renewcommand{\baselinestretch}{1.0}\normalsize
	\newpage
	
	\section{Methods and materials}
	The quadruple-quantum-dot device was fabricated on a $\mathrm{GaAs/Al_{0.307}Ga_{0.693}As}$ heterostructure grown by molecular-beam epitaxy. The metallic Ti/Au surface gates were patterned with electron-beam lithography. Quantum dots were formed in the 90-nm-deep two-dimensional electron gas having an electron density of $\mathrm{2.2 \cdot 10^{11}\ cm^{-2}}$ and a mobility of $\mathrm{3.4 \cdot 10^{6}\ cm^{2} V^{-1} s^{-1}}$ (measured at 1.3 K).
	The device was cooled down to a base temperature of $\sim$22~mK
	in an Oxford Triton 400 dilution refrigerator. Positive voltages were applied on the gates (ranging between 100 and 400~mV) while cooling down the device to reduce the magnitude of charge noise~\cite{Long2006}.
	
	Homebuilt bias-tees ($RC$=470~ms) were connected to gates P1, P3, and P4 for simultaneous application of a d.c.~voltage bias and high-frequency signals. Voltage pulses to the gates were applied using a Tektronix AWG5014. A microwave generator (HP83650A) was additionally connected to P3 via a homemade bias-tee at room temperature. For reading out signals on both sensing dots (SDs), we used a combination of radio-frequency (RF) reflectometry and frequency multiplexing. Frequencies were matched to LC circuits containing microfabricated superconducting spiral inductors. SD1 was connected to 3.1~$\mu$H and measured at 104.3~MHz while SD2 was connected to 4.6~$\mu$H and measured at 83.4~MHz. The reflected signals were amplified through a Weinreb CITLF2 amplifier at 4~K and subsequently demodulated at room temperature. Sensor signals were acquired through a field-programmable gate array (FPGA DE0-Nano Terasic), which applied threshold analyses to count spin detection events.
	
	\section{Detailed information of the applied pulse sequence}
\begin{figure}[!h]
		\centering
		\includegraphics[width=0.6\textwidth]{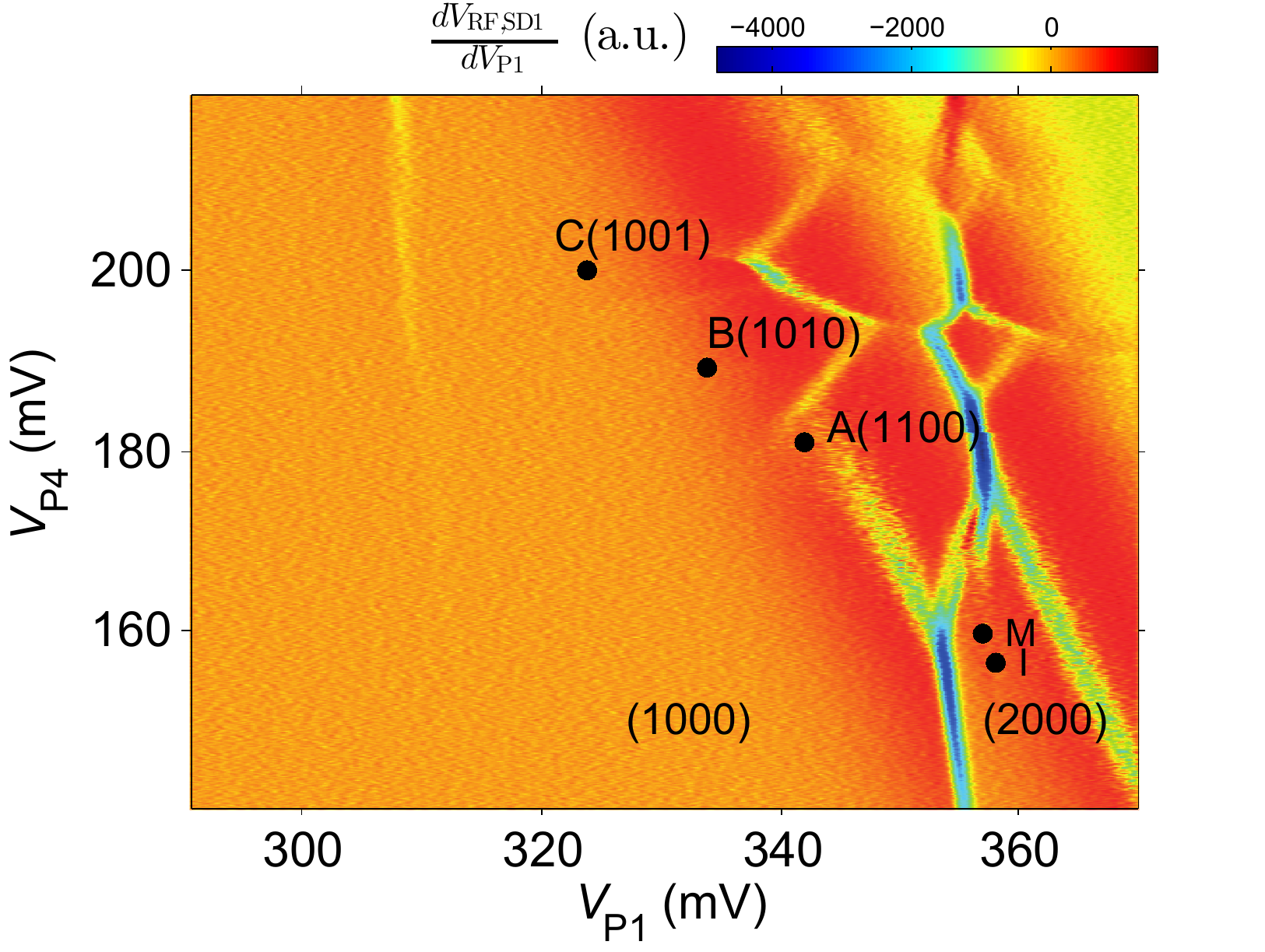}
		\caption{Charge stability diagram prepared for a coherent shuttle sequence. During the pulse sequence, the d.c.~voltages applied to the gates are set at the centre of the inter-dot transition line of (2000)-(1100). The gate voltages are next pulsed towards the depicted points I, A, B, C, and M. Although faint, inter-dot charge transition lines separate the points A, B, and C, which help distinguish each charge state. Added details for the pulse sequences are explained in Table S~\ref{tab:details_pulse_sequence_coherent_shuttle}.}
		\label{fig:honeycombs_detailed_pulse_sequence} 
	\end{figure}

	In this section, we show the charge stability diagram used for configuring the pulse measurements and give detailed information on the applied pulse sequences. Fig.~S\ref{fig:honeycombs_detailed_pulse_sequence} shows a similar charge stability diagram as Fig.~1c from the main text. To achieve a high spin read-out fidelity of the PSB measurement occurring between dot~1 and 2, SD1 is tuned to be most sensitive at the measurement point, M. As a consequence, some of the charge transition lines (e.g. those of dot 4) are not always clearly visible in Fig.~S\ref{fig:honeycombs_detailed_pulse_sequence}. The points A, B, C, M and I in the figure represent the exact gate voltages applied in the single-shot measurements. Table~S~\ref{tab:details_pulse_sequence_coherent_shuttle} gives an explanation of each pulse stage including the details on its duration.
	
	\begin{table}[h!]
		\centering
		\begin{tabular}{ |l | p{13cm} |}
			\hline
			\textbf{Stage(s)} & \textbf{Purpose and details} \\ \hline
			I (initialize) & Initialization stage. Wait 300~$\mu$s for relaxation to $S$(2000) spin state. \\ \hline
			A, B, or C & Manipulation stage. Pulse to shuttle through multiple inter-dot transitions, ending at either A(1100), B(1010), or C(1001) charge states, respectively. At the end-point A,B, or C we keep the pulse amplitude fixed for a varying amount of time (up to tens of nanoseconds). This allows coherent evolution between the  $S$ and $T_0$ states due to the charge-state dependent Zeeman energy difference $\Delta E_z$. \\ \hline
			M (read-out) & Read-out stage. Pulsing back into the (2000) region, spin-singlet and triplet states are read out via Pauli spin blockade (PSB). We wait 3.5~$\mu$s for the RF signal to reach its final value (the bandwidth was set to 500~kHz with a low-pass filter) and then start integrating the SD signal for 5~$\mu$s. (see Sec.~\ref{secS:PSB_read-out_fidelity} for fidelity calculations) \\ \hline
			Compensation stage~~ & At the end of the pulse sequence we add a compensation stage of 300~$\mu$s with voltages chosen to ensure that the total d.c.~component of the pulse sequence is 0. This prevents unwanted offsets of the dot levels due to the	bias tees. \\ \hline
		\end{tabular}
		\caption{Detailed explanation of the applied pulse sequence for the single-shot coherent shuttle measurements. Stage labels correspond to the points depicted in Fig.~S\ref{fig:honeycombs_detailed_pulse_sequence}. These stages are sequentially executed and repeated for 10$^3$ to 10$^4$ times. The integrated signal at the read-out stage is processed in the FPGA using threshold detection and the binary result is output from the FPGA.}
		\label{tab:details_pulse_sequence_coherent_shuttle}
	\end{table}
	
	\section{Inter-dot tunnel couplings}
\begin{figure}[!h]
		\centering
		\includegraphics[width=1.0\textwidth]{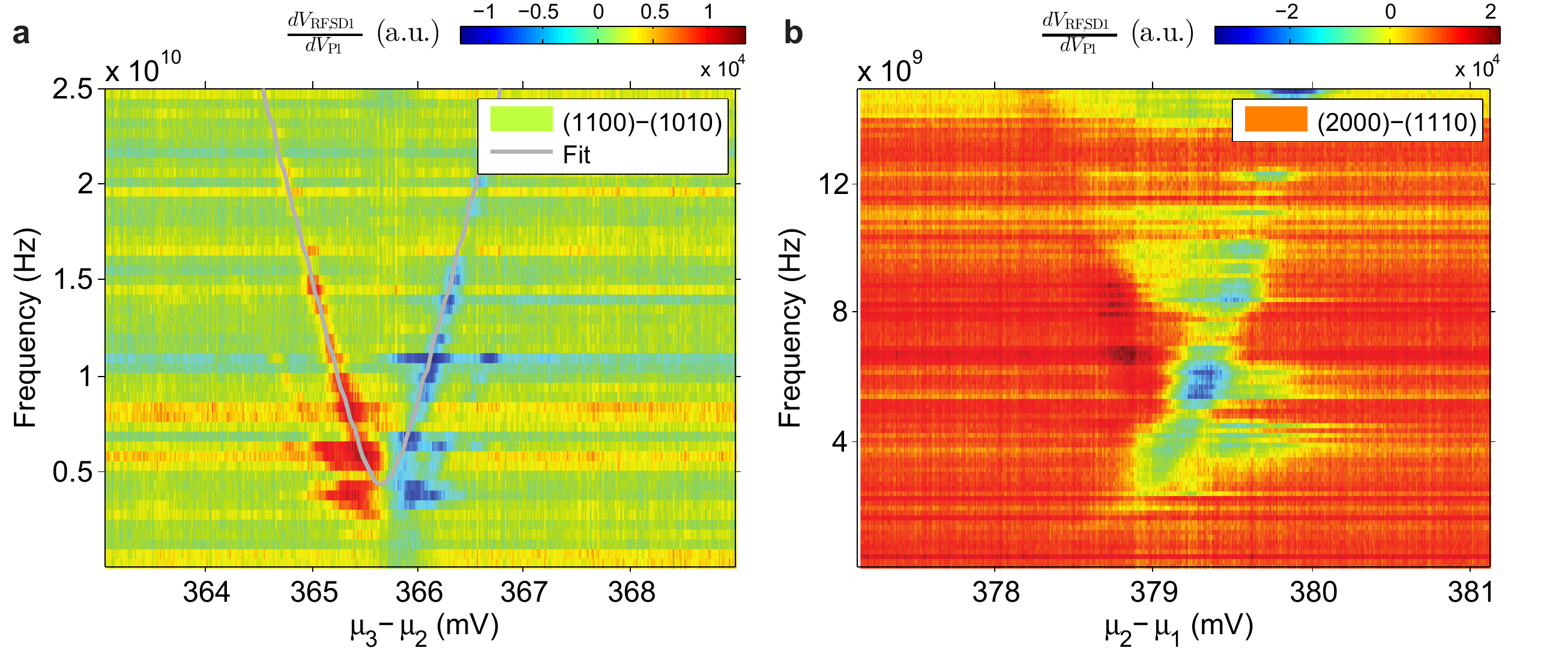}
		\caption{Photon-assisted tunnelling (PAT) measurements. {\bf a} PAT measurement across the inter-dot transition between the (1100) and (1010) charge states. The charge sensor signal shows a peak that forms a v-shaped curve in the detuning versus frequency diagram. The minimum frequency reached in the v-shape corresponds to the energy of the anti-crossing, $2t$. {\bf b} Similar to {\bf a} but on the (2000) and (1100) transition. The two-particle tunnelling process gives two-spin dependent transitions. The coupling strength is roughly tuned by estimating the curvature of the visible resonance lines.}
		\label{fig:PAT_measurement} 
	\end{figure}

	The three inter-dot tunnel couplings, $t_{12}$, $t_{23}$, and $t_{34}$, are suitably tuned to perform the coherent shuttling experiments. We aim for tunnel couplings of 2$\sim$3~GHz, for which adiabatic charge shuttling should be possible. The tunnel couplings were measured using photon-assisted-tunnelling (PAT)~\cite{Oosterkamp1998a}. An example for a single particle tunnelling measurement is shown in  Fig.~S\ref{fig:PAT_measurement}a for estimating the coupling between (1100) and (1010). The charge sensor response is monitored while sweeping across the inter-dot transition and applying a continuous microwave excitation to gate P3. The main resonance peaks are fitted to $\sqrt{(V-V_{\mathrm{offset}})^2+4t^{2}}$ by eye where $t$ is the tunnel coupling and $V$ is the detuning energy between the two relevant dots. Fig.~S\ref{fig:PAT_measurement}b shows the PAT measurement at the boundary between (2000) and (1100). At this inter-dot transition, PAT measurements show multiple transition lines of different curvature, related to the two-spin states that are involved in the tunnelling process (see Supplementary information of Ref.~\citenum{Baart2016}). It was difficult to accurately extract the tunnel coupling from this figure. A better estimate of the tunnel coupling at this transition and a comparison with experiment is made in the simulation section later below.
	
	\section{Details on calculation of the read-out fidelities}
	\label{secS:PSB_read-out_fidelity}
	In this section, we discuss how we obtain the read-out fidelities from the calibration data. Since our read-out method is similar to what has been used in previous works, we follow the approach discussed in Ref.~\citenum{Barthel2009Phys.Rev.Lett.}.
	
	\subsection{Analytic expressions for the fidelity}
	Singlet and triplet spin states are distinguished using spin-to-charge conversion at the measurement position. For a spin-singlet state, the system can return to a doubly occupied charge state (2000), whereas for a spin-triplet state, the system will remain in the (1100) charge state due to the Pauli exclusion principle. The fidelity of spin read-out will hence be influenced by the quality of the charge detection. Additionally, spin relaxation of the (excited) spin-triplet state during the measurement time will degrade the triplet read-out fidelity. 
	Using an FPGA, the RF read-out signal is integrated for a time $\tau_M$ and the result $V_{\rm rf}$ is compared to a certain threshold $V_{\rm T}$. Only the bit value expressing whether the signal goes above or stays below the threshold is recorded. We will now use the analytical expressions of Ref.~\citenum{Barthel2009Phys.Rev.Lett.} as the starting point for calculating the read-out fidelities for this experiment.
	
	We first discuss the statistics of the single-shot outcomes to calculate the expected read-out fidelities for each spin state. We find that the signal for (2000) and (1100) follow a noise-broadened Gaussian distribution, with equal broadening $\sigma$ and peak positions located at output voltages $V_{\rm rf}^S$ and $V_{\rm rf}^T$, respectively. If we focus on the probability density of measuring a given spin state for an output signal $V_{\rm rf}$, the probability density for measuring a spin-singlet state, $n_S$, is described by,
	\begin{equation}
		n_S(V_{\rm rf})=\left(1-\left< P_T \right>\right)e^{-\left[\left(V_{\rm rf}-V_{\rm rf}^S\right)^2/2\sigma ^2\right]}\frac{1}{\sqrt{2\pi}\sigma},
	\end{equation}
	where $\left<P_T\right>$ is the probability of measuring a triplet state in the total measurement run. For triplet states, we also need to take a spin relaxation process into account during the signal integration time. The probability density of the triplet states, $n_T$, is then given by,
	\begin{equation}
		n_T\left(V_{\rm rf}\right)=e^{-\tau _M/T_1}\left<P_T\right>e^{-\left[\left(V_{\rm rf}-V_{\rm rf}^T\right)^2/2\sigma^2\right]}\frac{1}{\sqrt{2\pi}\sigma}+\int_{V_{\rm rf}^S}^{V_{\rm rf}^T}{\frac{\tau_M}{T_1}\frac{\left<P_T\right>}{\Delta V_{\rm rf}}e^{-\left[\left(V-V_{\rm rf}^S\right)/\Delta V_{\rm rf}\right]\left(\tau_M/T_1\right)}e^{-\left[\left(V_{\rm rf}-V\right)^2/2\sigma^2\right]}\frac{dV}{\sqrt{2\pi}\sigma}},
	\end{equation}
	where $\Delta V_{\rm rf}=V_{\rm rf}^T-V_{\rm rf}^S$ is the signal contrast between the two states and $T_1$ the relaxation time from a triplet state to a singlet state.
	
	The read-out fidelity of a singlet (triplet) state, $F_S$ ($F_T$) can be found after calculating the error rates from the densities $n_S$ and $n_T$. Each of the fidelities are expressed in the following form by respectively inputting $P_T=0$ and $P_T=1$,
	\begin{equation}
		F_S=1-\int_{V_{\rm T}}^{\infty}n_S(V)dV, \,\,\,\,\,\,\, F_T=1-\int_{-\infty}^{V_{\rm T}}n_T(V)dV.
		\label{eqS:fidelity_definition}
	\end{equation}
	The integrals in Eq.~\ref{eqS:fidelity_definition} are the probabilities of misidentifying a singlet as a triplet and vice versa that will depend on the threshold value $V_{\rm T}$. One could set an optimized $V_{\rm T}$ and $\tau _M $ to minimize the error rates for a specific measurement, however, here we took the following procedure to simplify the threshold determination and applied this to every single-shot measurement described in the main text.
	
	\subsection{Parameter determination and estimated read-out fidelities}
	After fixing the integration time, we took histograms to determine the optimal threshold value and measured $T_1$ from single-shot measurements. These parameters allow us to calculate the fidelities of our read-out protocol.
	
	\begin{figure}[!h]
		\centering
		\includegraphics[width=1.0\textwidth]{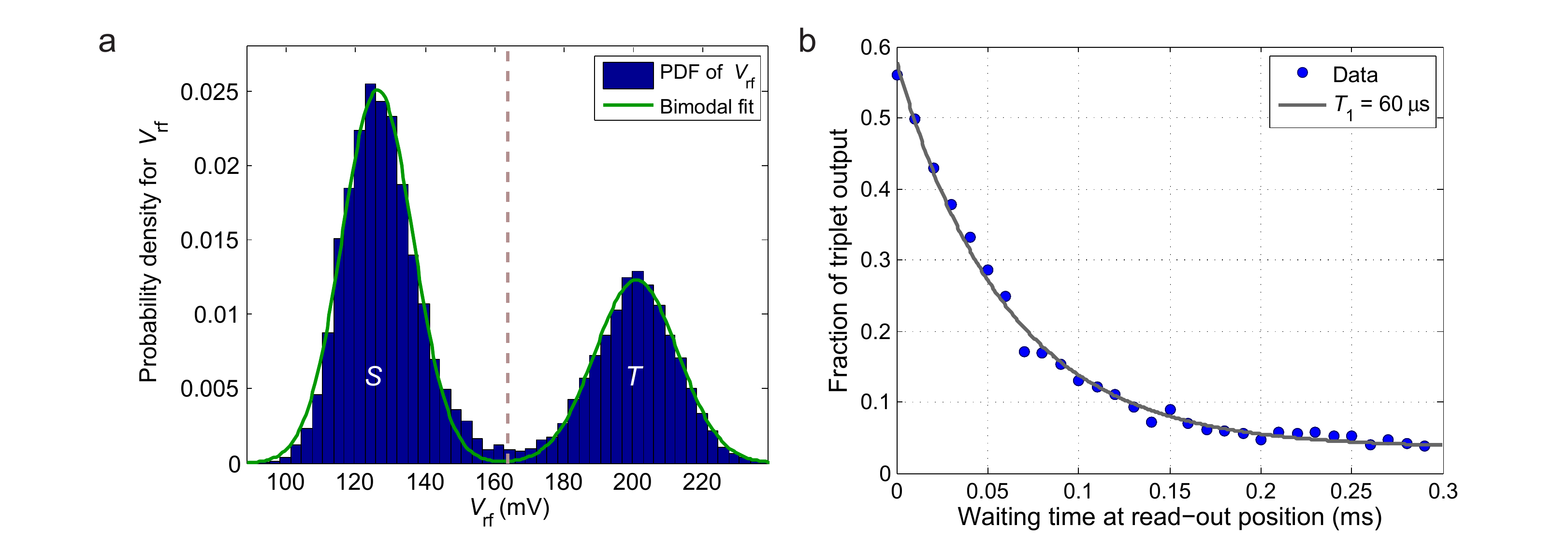}
		\caption{Measurement results for calibrating the single-shot PSB measurement. {\bf a} Example histogram of RF-signal. After initializing to a mixture of $S$ and $T_0$ states, the lower (higher) $V_{\rm rf}$ peak is attributed to a singlet (triplet) outcome. Based on Gaussian fits to the peaks we determine the threshold voltage, $V_{\rm T}=162$~mV for this example. {\bf b} Triplet state relaxation as a function of waiting time at the read-out stage. Fitting to an exponential decay gives $T_1=60~\mu$s and an offset of 3.6~\%. }
		\label{figS:FPGA_outcome_PSB_calibration}
	\end{figure}
	
	Figure~S\ref{figS:FPGA_outcome_PSB_calibration}a shows an example histogram of the measured time averaged signal fitted to a bimodal distribution of normal Gaussians. For this histogram we first randomized the spin by waiting in the (1001) charge state for 1~$\mu$s. After pulsing the voltages to the measurement point in (2000), we first let the voltage response settle for 3.5~$\mu$s. Then the output signal was integrated for $\tau_M=5~\mu$s. Such sequences were repeated 10,000 times. Here, the fitting gave $\Delta V_{\rm rf}=74.5$~mV, and $\sigma=11.1$~mV. We chose $V_{\rm T}$ to take a value centred between the two peaks. For the calibration procedure, we averaged only over of order 1,000 points, in order to reduce its duration. This may 
	give variations in $V_{\rm T}$ of a few mV.
	
	After calibrating the threshold voltage, we can perform single-shot measurements such as the $T_1$ measurement shown in Fig.~S\ref{figS:FPGA_outcome_PSB_calibration}b. Fitting to an exponential decay results in a relaxation time $T_1=60\pm 4$~$\mu$s. The $T_1$ values are stable for at least a few days unless we change the gate voltage settings. The offset in this fit represents the read-out error for a singlet state. The value of the fitted offset, 3.6$\pm 0.8$~\%, is larger than expected from statistics based on the parameters shown up to now, or from thermal excitation. Here, for measurements containing longer voltage pulses, there may be a drift in the applied voltage due to the finite bandwidth of the bias tee. This may cause an overall shift of -17~mV in $V_{\rm T}$ resulting in the displayed offset.
	
	Finally we discuss the fidelities calculated from the given formula. Using the obtained fitting parameters and ideally applying our procedure, we expect fidelities of $F_S=99.96\pm 0.03$~\% and $F_T=95.9~\pm 0.3$~\%. For the worst case scenario we take the condition where a voltage drift may cause $V_{\rm T}$ to shift by -17~mV. Then we obtain $F_S=96.6\pm 0.8$~\% and $F_T=97.7~\pm 0.1$~\%. This allows us to conclude that our read-out fidelities are above 95\%. The estimated read-out fidelities are high enough to exclude the possibility of spin read-out errors being the main cause of the reduced contrasts in the $S$-$T_0$ oscillations shown in the main text.
	
	\section{Details on the Zeeman splitting measurements}
	EDSR is performed to determine the local Zeeman energy in each dot. We use adiabatic passage to flip the single-spin states \cite{Shafiei2013a}. The microwave pulses are applied on a common gate, P3, to drive transitions of the spins for every dot. The read-out sequence is now operated in an energy-selective spin-read-out mode to measure the single spins, the same way as in Ref.~\citenum{Baart2015}.
	
	\begin{figure}[!h]
		\centering
		\includegraphics[width=0.6\textwidth]{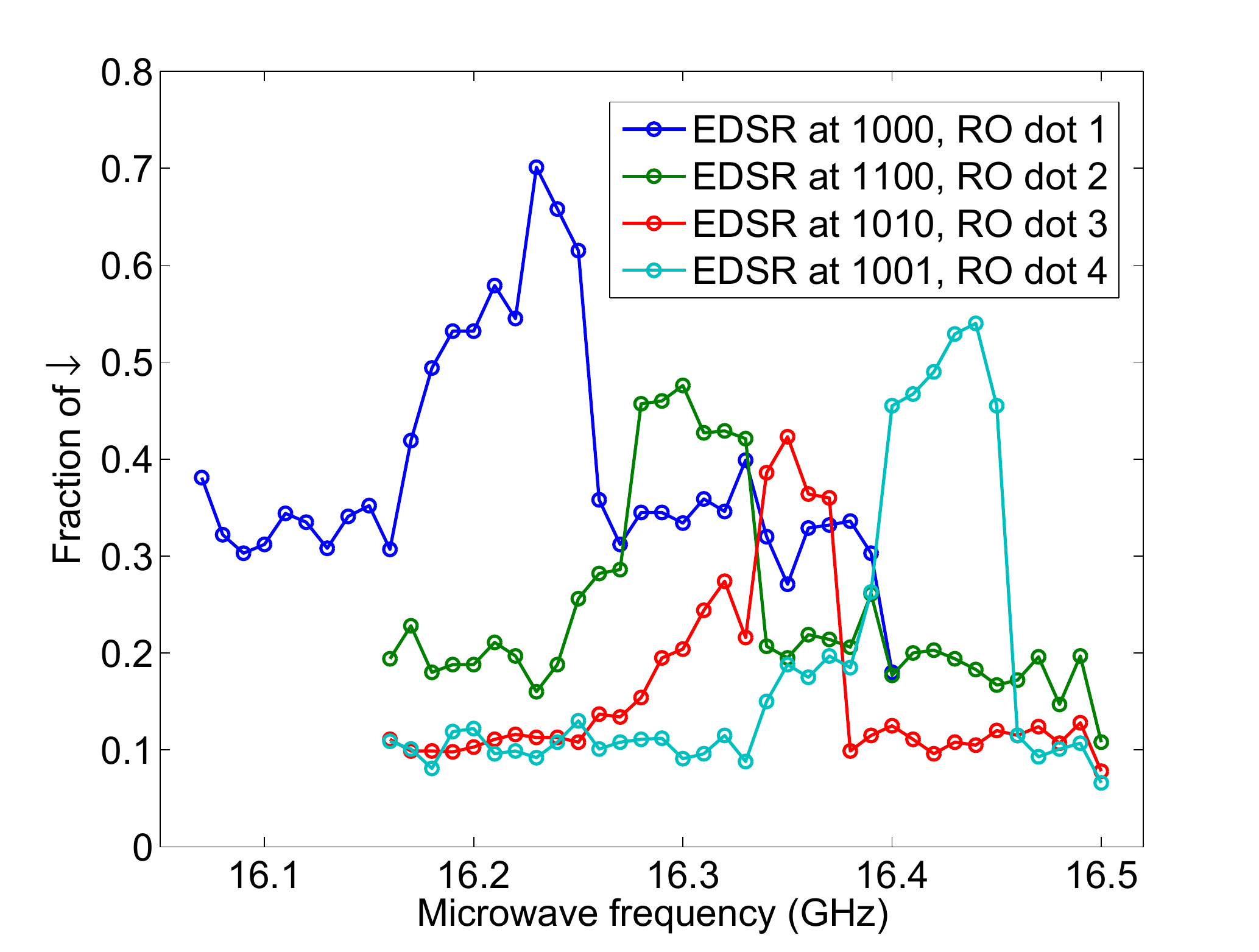}
		\caption{An example of electric-dipole spin resonances at 2.7 T. A single electron is first initialized to spin up. Next, frequency-modulated microwave bursts are applied whilst the electrons are residing in the denoted charge states. Read-out is performed for each of the single spins through one of the reservoirs. The spin-state of dot 1 is read out via the left reservoir. The spin-state of  dots 2, 3, and 4 is read out via the right reservoir, using the single-spin CCD concept as described in Ref.~\citenum{Baart2015}.}
		\label{figS:edsr_spectra}
	\end{figure}
	
	Fig.~S\ref{figS:edsr_spectra} shows an example of the EDSR spectra for each dot at 2.7~T. Similar measurements are performed at magnetic fields between 2.7 and 4.7~T as in the points presented in Fig.~4a in the main text. The resonance peaks have an uncertainty coming from two factors: (1) the measurement resolution and (2) quasi-static nuclear spin field fluctuations. The resonance peak is first broadened by the frequency modulation needed for adiabatic passage (44 MHz modulation in 400 $\mu$s). We extracted the highest frequency from each trace where the signal is visible. To compensate for the frequency modulation, we subtract 22 MHz from the extracted frequency. The uncertainty of the extraction mainly comes from the measurement resolution of $\pm$5~MHz. The second factor relates to the time taken to average the EDSR signals. Each point in the trace is averaged over roughly 10 seconds. This measurement time is shorter than the correlation time (of the order of 100 seconds~\cite{Bluhm2010,Delbecq2016}) of the nuclear spins, meaning that the electron spin feels roughly a static magnetic field during a single measurement point. However, each frequency trace takes a few minutes to complete, therefore, the extracted frequency can depend on the nuclear spin configuration at the exact timing when the point is measured. We expect the resonance peak position to vary by roughly $\pm$15~MHz due to this nuclear spin effect.
	
	Furthermore, we observed that the resonance frequencies shift with gate voltage. In this device, the resonance frequencies shifted up to 300~MHz at 2.7~T with gate voltage variations on the order of 5 to 15 mV. Table \ref{tab:example_resonance_extra} lists an example of the gate voltages and corresponding spin resonance frequencies for dot~1 and dot~4. For the EDSR measurements presented in the main text, the resonance frequency of dot~1 was measured in the (1000) charge region, while the coherent oscillations were performed in the charge states (1100), (1010), and (1001). The roughly 50~MHz difference in $\Delta E_z^{i,j}$ observed between the EDSR and coherent shuttling measurements may be caused by the differences in the gate voltages used.
	
	\begin{table}[h!]
		\centering
		\begin{tabular}{ | c | c | c |}
			\hline
			\textbf{Gate [mV]} & \textbf{Condition A} & \textbf{Condition B} \\ \hline
			D1 & -395.0 & -390.0 \\ \hline
			D2 & -480.0 & -475.0 \\ \hline
			D3 & -261.0 & -256.0 \\ \hline
			P1 & 255.4 & 253.8 \\ \hline
			P2 & 78.1 & 74.3 \\ \hline
			P3 & -89.2 & -98.7 \\ \hline
			P4 & 40.3 & 29.7 \\ \hline
			SD1b & -247.5 & -246.5 \\ \hline
			SD2b & -34.0 & -30.0 \\ \hline
			\hline
			$E_{z,1}/\hbar$ [GHz] & 15.48 & 15.78 \\ \hline
			$E_{z,4}/\hbar$ [GHz] & 15.52 & 15.75 \\ \hline
		\end{tabular}
		\caption{Resonance frequencies of dot~1 and dot~4 for two different gate voltage conditions. The gate voltages shown are the ones that are varied. Other gate voltages are identical between the two measurements. The observed variation in the resonance frequencies is larger than the random fluctuations attributed to the limited measurement resolution and random nuclear spins.}
		\label{tab:example_resonance_extra}
	\end{table}
	
	\section{Simulation of the coherent shuttling}
	Below we describe how we obtained the contrast values in Figs.~4b,c,d of the main text. Later on we show simulation results including the read-out pulse.
	
	\subsection{Simulation model}
	Our experimental condition is modelled with the following Hamiltonian, defined in the $S$(2000), $S$(1100), $T_0$(1100), $S$(1010), $T_0$(1010), $S$(1001), and $T_0$(1001) basis.
	\[ H = \left( \begin{array}{ccccccc}
	\mu_1	&\sqrt{2}t_{12}	&0	&0		&0		&0		&0\\
	\sqrt{2}t_{12}	&\mu_2	&\Delta E_{z,21}/2	&t_{23}	&0	&0	&0\\
	0	&\Delta E_{z,21}/2	&\mu_2	&0	&t_{23}	&0	&0\\
	0	&t_{23}	&0	&\mu_3	&\Delta E_{z,31}/2&t_{34}	&0\\
	0	&0	&t_{23}	&\Delta E_{z,31}/2	&\mu_3	&0	&t_{34}\\
	0	&0	&0		& t_{34}	&0	&\mu_4	&\Delta E_{z,41}/2\\
	0	&0	&0		&0	& t_{34}	&\Delta E_{z,41}/2	&\mu_4\\
	\end{array} \right) \]
	The parameters are defined as follows. $\mu_i$ is the potential of dot $i$. $\Delta E_{z,ji}=E_{z,i}-E_{z,j}$ is the Zeeman energy difference between dot $i$ and $j$, extracted from the fits to the oscillation in Figs.~3a,b,c in the main text. $t_{ij}$ is the single-particle tunnel coupling between dot $i$ and $j$. The tunnel couplings were fixed to the following values $t_{12}=1.2$~GHz, $t_{23}=2.6$~GHz, and $t_{34}=4.3$~GHz, as also noted in the main text, throughout this section unless they were explicitly varied.
	
\begin{figure}[!h]
		\centering
		\includegraphics[width=1.0\textwidth]{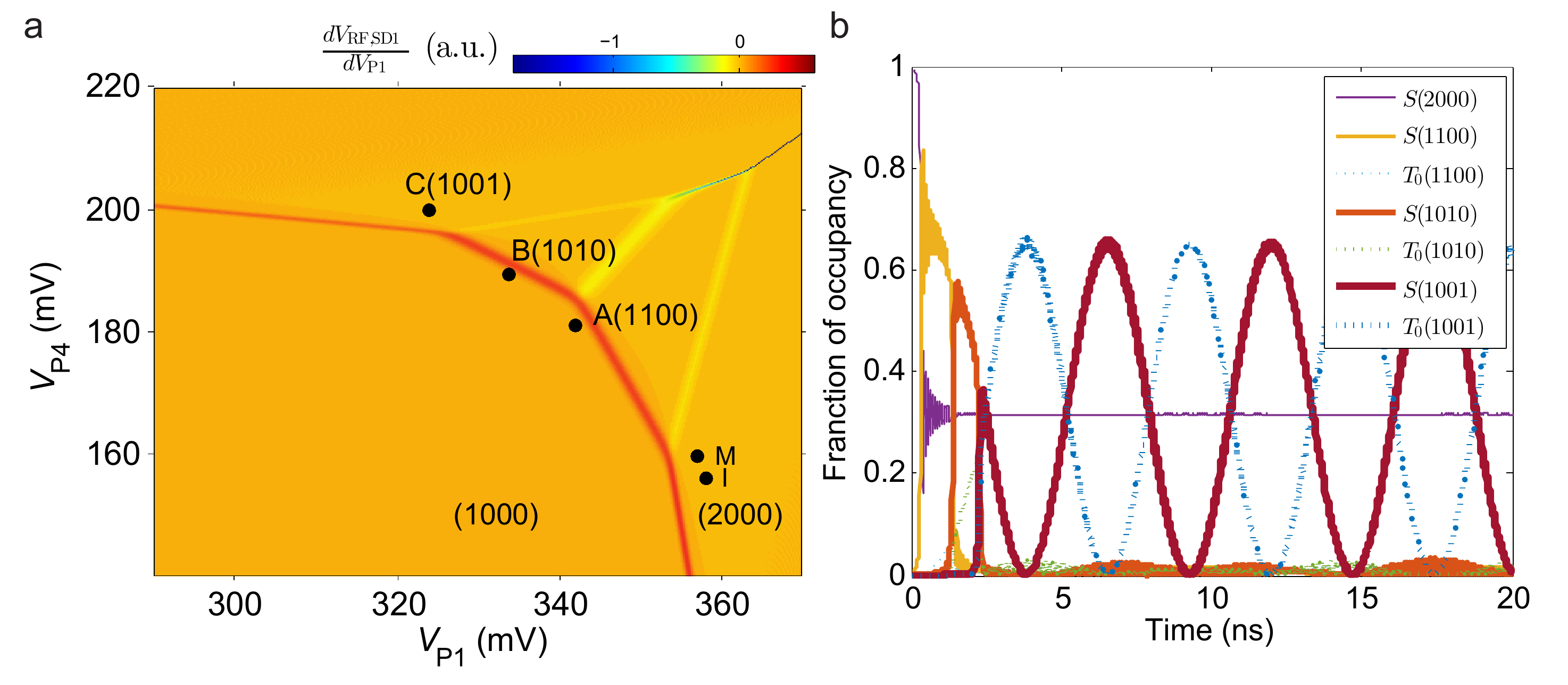}
		\caption{Simulation of coherent shuttling. {\bf a} Numerical calculation of the charge sensor response for the voltages used to measure Fig.~S\ref{fig:honeycombs_detailed_pulse_sequence}. The relevant two-electron charge state regions used for the coherent shuttling experiment and that of the (1000) state are reconstructed. {\bf b} Simulation of the state evolution while pulsing from point I in (2000) to point C in (1001). The contrast of the $S$-$T_0$ oscillation plotted in Fig. 4d of the main text is calculated here from the oscillation amplitude after 2.5~ns. The inter-dot tunnel couplings are set to $t_{12}=1.2$~GHz, $t_{23}=2.6$~GHz, and $t_{34}=4.3$~GHz.}
		\label{figS:simulation_result}
	\end{figure}

	Figure~S\ref{figS:simulation_result}a shows an example of the reproduced charge-stability diagram based on this Hamiltonian. The chemical potentials, $\mu_1, \mu_2, \mu_3,$ and $\mu_4$~[$\mu$eV], were converted from gate voltages, $V_{\rm  P1}$ and $V_{\rm  P4}$~[mV], at each point using the following conversion formula.
	\[ \left( \begin{array}{c}
	\mu_1\\
	\mu_2\\
	\mu_3\\
	\mu_4\\
	\end{array} \right)
	= \left( \begin{array}{cccc}
	-49.0	&-6.15\\
	-25.3	&-11.1\\
	-13.1	&-20.0\\
	-6.33	&-52.0\\
	\end{array} \right) \left( \begin{array}{c}
	V_{\rm P1}\\
	V_{\rm P4}\\
	\end{array} \right)
	+ \left( \begin{array}{c}
	18300\\
	10700\\
	8110\\
	12300\\
	\end{array} \right) \]
	The 8 matrix elements in this convention were calculated from measured cross capacitances and a lever arm between gate voltage and energy. Cross capacitances between the gates and the dots are measured from the slopes of the 4 charging lines, and the 3 inter-dot transition lines observed in Fig.~S\ref{fig:honeycombs_detailed_pulse_sequence}. The lever-arm, which converts gate voltage to dot energy, was extracted from a spectroscopy measurement using photon-assisted tunnelling~\cite{Braakman2013a}. For this, we used the effect of $V_{\rm P1}$ on the detuning energy, $\mu_4-\mu_3$, which was measured to be 32.6~$\mu$eV/mV. Additionally, an energy offset is included as the energy where the (1000) charge state aligns to the fermi-energy. We calculated the 4 offsets from the 3 triple-point coordinates and defining those coordinates as zero energy. Note that the following shuttling simulations are valid regardless of this offset. We have included the offsets to get good correspondence between the experimental and simulated charge diagrams.
	
	We included a time dependent gate voltage function to incorporate the finite risetime of the pulse. A low pass filter (SLP-300+ Mini-Circuits) was added after the AWG output channel that shaped the voltage ramp to an approximate linear shape from 5~\% to 95~\% of the signal. The important inter-dot transitions sit inside this region of the pulse. Therefore we modelled the total pulse shape to be linear with a total rise time of 2.50~ns.
	
	The evolution of the spin states during shuttling is simulated by discretizing the time dependent operations with steps of $\Delta t=0.01$~ns. The initial state is set to the ground state at point I in the diagram. Then we calculate the time evolution at each point of the gate voltage pulse using the solution of the time-dependent Schr\"odinger equation,
	\[
	\Psi (t+\Delta t) = {\rm exp}^{-i\frac{H(t)\Delta t}{\hbar}}\Psi (t).
	\]
	
	\subsection{Simulation results}
	Figure~S\ref{figS:simulation_result}b shows an example of a simulation result of a spin shuttle from point I to point C. The initial spin state at $t=0$ was set to the ground state in the (2000) region. As the gate voltages were ramped, most of the $S$(2000) state first propagated to $S$(1100). However, a fraction of 0.32 remained in $S$(2000). This remainder is the non-adiabatic portion of the charge transfer at the first inter-dot transition. The portion that was transferred, starts a $S$-$T_0$ oscillation while the charge makes a transition from (1100), (1010), to (1001). When the ramp ends in point C, most of the population ends up in the (1001) charge state. Only relatively low fractions of (1100) and (1010) charge states remain in the end, indicating good charge adiabaticity from (1100) to (1010) and (1001). This can be explained by the higher tunnel couplings for the latter transitions. The oscillation frequency depends on the final charge state. For a final charge configuration (1001), it is given by $\Delta E_{z,41}$. The oscillation contrast was estimated by extracting the difference between the maximum and the minimum fraction of the $T_0$ state. This procedure was repeated for every tunnel coupling value in Fig.~4 in the main text. 
	
	\begin{figure}[!h]
		\centering
		\includegraphics[width=1.0\textwidth]{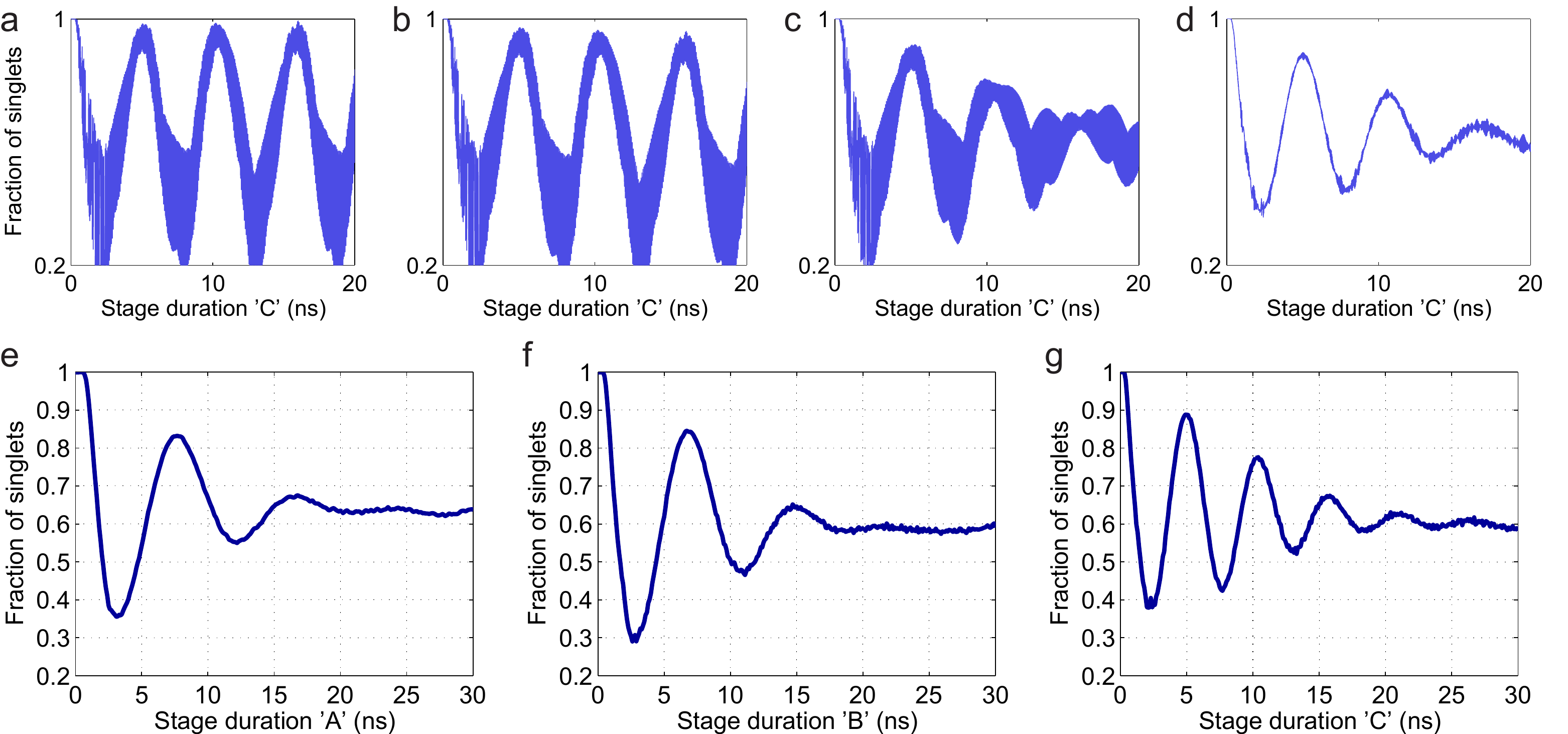}
		\caption{Simulation procedure for including the read-out sequence. {\bf a} Calculated singlet probabilities after applying the measurement operations after each duration time. {\bf b} Including charge relaxation throughout the shuttling experiment, with a time constant of 50~ns. {\bf c} Averaging over 100 random nuclear spin distributions, with a standard deviation of 15~MHz. {\bf d} Including, for each nuclear spin distribution, a random contribution to the dot potentials with standard deviation of 4~$\mu$eV. {\bf e,f,g} Calculated singlet probabilities after both the pulses from point I to points A, B, or C and then back to M. Averaged over 1000 nuclear spin and dot potential configurations.}
		\label{figS:simulation_result_ oscillations}
	\end{figure}
	
	To further compare with the detailed experimental observations, including the damping, offsets and phase shifts seen in the $S$-$T_0$ oscillation measurements, we simulated the pulse sequence including the read-out pulse (see Fig.~S\ref{figS:simulation_result_ oscillations}a). The read-out pulse is a voltage ramp to the measurement point, M, after the waiting stage. This voltage ramp is implemented in the simulation in a similar way as the voltage ramp for the forward pulse from point I to point A, B, or C, by solving the time dependent Schr\"odinger equation. After reaching point M, the final state is recorded. At this point in time, experimentally the potential is kept constant for signal integration on the order of micro seconds, which we assumed to be long enough for the charge states to relax to the measurement basis. Therefore, we included in the simulation complete charge relaxation at the final stage. At the same time, we incorporated inelastic tunnelling during the forward shuttling and waiting time. We also accounted for random nuclear spin fluctuations and charge noise by averaging over multiple Zeeman splitting values and dot potentials. The implementation of these processes is explained below.
	
	Charge relaxation during the read-out phase is modelled by transferring the higher energy state fractions to the lower energy states. Presumably, the relevant lower energy states are the states $S$(2000) and $T_0$(1100), which constitute the measurement basis states during the PSB measurement. Relaxation to these states from the higher energy states occurs through inelastic charge tunnelling. An inelastic tunnelling rate of (6~ns)$^{-1}$ is measured on a GaAs based double-dot structure, when the detuning energy between the two dots is 30~$\mu$eV and the tunnel coupling is 9~$\mu$eV~\cite{Hayashi2003}. Our tunnel couplings are on the same order but the detuning energies between the relevant charge states are more than ten times larger (minimum of 315~$\mu$eV). In the simulation, we therefore set the charge relaxation rate to 50~ns assuming a 1/$\varepsilon$ dependence~\cite{Fujisawa2000}.
	
	During the read-out stage, the higher energy eigenstates corresponding to the (1010) and (1001) configuration are in the $\uparrow\downarrow$ basis. Therefore, the electron spins will continue to evolve between $S$ and $T_0$ due to the Zeeman energy difference. The oscillation rate of $S$-$T_0$ is on the order of 100~MHz, which is fast compared to the charge relaxation rates. This results in a random phase between $S$ and $T_0$ after relaxation. In the model, we assumed for simplicity total dephasing of the spins and that relaxation occurs equally to $S$(2000) and $T_0$(1100). For the higher energy states corresponding to the (1100) configuration, we take the $S$(1100) and $T_0$(1100) state, which are separated by the exchange interaction. Here we assume that $S$(1100) selectively relaxes to $S$(2000) .
	
	Charge relaxation (inelastic tunnelling) is also taken into account for the earlier parts of the experiment, upon shuttling and while at points A, B or C. The main effect is to cause an offset in the observed $S$-$T_0$ oscillation away from 0.5 (the oscillations saturate around 0.6 in Figs.~S\ref{figS:simulation_result_ oscillations}e,f,g). The fraction of the initial $S$(2000) state that is not adiabatically transferred to (1100), (1010) or (1001) after the pulse from I to A, B or C, remains in $S$(2000), increasing the final $S$ fraction observed in the measurement stage. This is also seen experimentally as the singlet fraction saturation around 0.6 in Fig.~3 of the main text.
	
	Finally, the simulation averages over nuclear spin configurations and over detuning, to account for charge noise. Accounting for random nuclear spin distributions, we assumed a standard deviation of 15~MHz ($\sim$2.5~mT) for the Zeeman splitting in each dot (see Fig.~S\ref{figS:simulation_result_ oscillations}c). Charge noise is incorporated by adding a random contribution of 4~$\mu$eV~\cite{Baart2016} to each dot potential. This has an effect of diminishing the charge interference effect during the forward and backward shuttling (see Fig.~S\ref{figS:simulation_result_ oscillations}d). Final results simulated for the three shuttling positions are presented in Figs.~S\ref{figS:simulation_result_ oscillations}e,f,g averaging over 1000 traces. In all the above calculations, no read-out errors are included. Overall, the final simulated traces resemble the experimental results in Fig.~3 of the main text and thus further strengthen our interpretation of the data in terms of electron shuttling while preserving the spin phase, whereby the measured fractions in the $S$-$T_0$ oscillation reflect the charge and spin adiabaticity conditions.

\end{document}